# Comparison of Source Coding Techniques for the Vehicle to Vehicle Communication

Subrahmanya Gunaga
B. E. Final Year : School of Electronics and
Communication Engineering
KLE Technological University
Hubballi, India

Rahul M. S.
B. E. Final Year : School of Electronics and
Communication Engineering
KLE Technological University
Hubballi, India

Varad Vinod Prabhu
B. E. Final Year : School of Electronics and
Communication Engineering
KLE Technological University
Hubballi, India

Akash Kulkarni
Research Assistant Professor : School of Electronics and
Communication Engineering
KLE Technological University
Hubballi, India

Nalini C. Iyer
Professor : School of Electronics and
Communication Engineering
KLE Technological University
Hubballi, India

*Abstract*—**Autonomous driving is gaining its importance due to the advancements in technology. With the intention of safety during human driving and with the longer-term aim to act as a communication enabler for autonomous driving, vehicle to vehicle communication is gaining its importance. In this paper, we discuss and compare various source coding techniques that can be used for vehicle to vehicle communication. We propose abbreviation-based and probability-based source coding methods for the vehicle to vehicle communication. We compare the proposed application-specific source coding methods with other techniques like Huffman, Arithmetic, and Lempel-Ziv-Welch coding. Experimental results show that the proposed probability-based source coding has better values of the compression ratio to the time required for all the messages considered.**

*Keywords— Huffman, Arithmetic, Lempel-Ziv-Welch, Abbreviation-based, Probability-based*

## I. INTRODUCTION

In this paper, we propose the method for source coding of safety messages for communicating among vehicles. Autonomous driving is the navigation process in which the vehicle can sense the surrounding environment and make decisions according to changing environmental conditions. Generally, the autonomous systems depend on sensors like cameras, radar, to sense the environment. But most of the sensors require a line of sight to detect the environmental changes. Thus, the vehicle to everything (V2X) is a technology used to share the information by communication. Similarly, vehicle to vehicle (V2V) communication is part of V2X where vehicles communicate among themselves to provide additional safety during driving. In recent years, the advancement in communication technologies vehicle to vehicle communication is gaining its importance. Vehicle to Vehicle communication is mainly helpful as it helps prevent accidents, the early intimation of emergencies, and applications related to autonomous vehicles.

The digital communication system's main steps at the transmitter are source coding, error correction coding, channel coding, and modulation. Similarly, at the receiver, the reverse operation is performed to get the information message that is sent. The steps involved in the receiver are demodulation, channel decoding, error correction decoding, and source decoding, as shown in Figure 1.

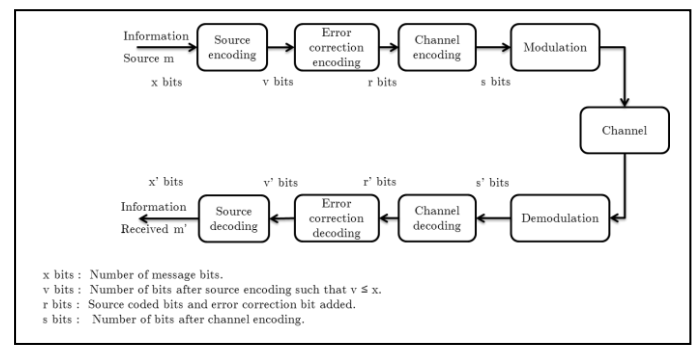

Fig. 1. General block diagram of a digital communication system

Source coding or data compression in digital communication is a technique where the symbols from the information source are mapped to a sequence of bits. The information can be recovered from the mapped sequence of bits. Based on the exactness of the retrieved information message from the course of bits, the source coding technique is broadly classified into two types of lossless source coding and lossy source coding, respectively. Since lossless source coding can recover the original information from the compressed sequence of bits, lossless source coding is more suitable for an automobile terminal communication system [1]. The general block diagram of lossless source coding is shown in Figure 2. The main aim of source coding is to remove the redundancy in the data to be transmitted. There are two broad categories of lossless source coding techniques, namely entropy-based and dictionary-based coding, respectively. The entropy-based coding technique is suitable if the entropy characteristics of the source data are known. In contrast, dictionary-based coding is ideal if the message to be transmitted is fixed and large.

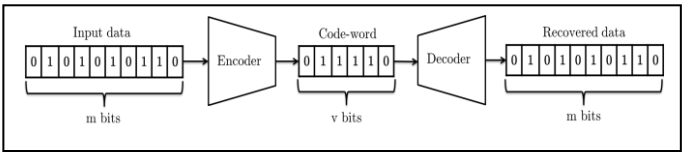

Fig. 2. General block diagram of lossless source coding technique where v ≤ m

## II. RELATED WORK

Several works have been done by researchers towards source coding or data compression. Authors of [2] discuss a new entropy-based method for multimedia coding. The proposed algorithm considers the occurrences of symbols in a given sequence based on which they are assigned a particular group and coded effectively using the suitable number of bits corresponding to that group. Experiments were performed on several data formats such as text, image, audio, and video. In the paper [3], the Double Huffman coding is used to compare Huffman coding's performance with the minimum variance Huffman coding. The author concludes that the double minimum variance Huffman coding is efficient. The compression is superior compared to the Huffman coding at the cost of increased computational and space complexity.

In paper [4], the authors identify a suitable source coding algorithm for compression of text, image, and audio data, respectively. For each of the data-formats, the source coding method chosen based on the compression ratios obtained. Authors in [5] compare the three commonly used lossless compression performance and efficiency based on time and space complexity. Experimental results show that the Huffman coding is suitable for compressing plain, whereas for image data LZW compression algorithm is suitable. For compressing hybrid text files like DOC, PDF, DOCX, and audio, Shannon-Fano coding is appropriate. Authors of the paper [6] discuss the performance of Huffman, Shannon, Shannon-Fano coding for binary, ternary and quaternary coding. Huffman coding algorithms are optimal for binary, ternary, and quaternary coding compared to the other two algorithms.

The authors of the paper [1] discuss various lossless compression methods and explain the advantages, disadvantages, principles of operation, and features of these algorithms. The article compares the compression ratio, CPU time, memory cost, and executable code size. LZ77, PPM, and BWT algorithms are suitable for electric vehicle communication.

## III. LOSSLESS SOURCE CODING TECHNIQUES

In this paper, we compare the performance of source coding methods such as Huffman, Arithmetic, and Lempel-Ziv-Welch coding with the proposed Abbreviation based coding, and Probability-based coding for the vehicle to vehicle communication of the safety messages. For comparison of all the algorithms, the 20 safety messages are considered as shown in Table II.

### A. Huffman coding

Huffman coding [9] is an entropy-based lossless source coding technique. It is a variable-length coding technique where each of the source symbols is mapped to variable-length codes. It works on the principle that more frequent the symbols occur, then it is represented by fewer bits. On the other hand, the less frequent source symbols are mapped to more number of bits. Huffman coding has the property that symbols are assigned to code-word resulting in prefix code or prefix-free code. Thus, the received code-word can be uniquely decoded by the decoder.

TABLE I: Probability of occurrence of English language characters

| S. L. No. | Character | Probability |
|---|---|---|
| 1 | a | 0.065174 |
| 2 | b | 0.012425 |
| 3 | c | 0.021734 |
| 4 | d | 0.034984 |
| 5 | e | 0.104144 |
| 6 | f | 0.019788 |
| 7 | g | 0.015861 |
| 8 | h | 0.049289 |
| 9 | i | 0.055809 |
| 10 | j | 0.000903 |
| 11 | k | 0.005053 |
| 12 | l | 0.033149 |
| 13 | m | 0.020212 |
| 14 | n | 0.056451 |
| 15 | o | 0.059630 |
| 16 | p | 0.013765 |
| 17 | q | 0.000861 |
| 18 | r | 0.049756 |
| 19 | s | 0.051576 |
| 20 | t | 0.072936 |
| 21 | u | 0.022513 |
| 22 | v | 0.008290 |
| 23 | w | 0.017129 |
| 24 | x | 0.001369 |
| 25 | y | 0.014598 |
| 26 | z | 0.000784 |
| 27 | Space | 0.191818 |

TABLE II: BASIC SAFETY MESSAGES FOR VEHICLE TO VEHICLE COMMUNICATION

| S. L. No. | Safety Messages | Abbreviations | Priority | Probability | Probability-based code |
|---|---|---|---|---|---|
| 1 | left turn ahead | LTA | $P_2$ | 50/1075 | 00111 |
| 2 | right turn ahead | RTA | $P_2$ | 50/1075 | 00110 |
| 3 | emergency ahead | EGA | $P_1$ | 100/1075 | 101 |
| 4 | emergency braking | EGB | $P_1$ | 100/1075 | 100 |
| 5 | brakes applied | BKA | $P_1$ | 100/1075 | 111 |
| 6 | lane change alert | LCA | $P_2$ | 50/1075 | 00001 |
| 7 | queue warning | QEW | $P_3$ | 25/1075 | 001001 |
| 8 | hump warning | HMW | $P_3$ | 25/1075 | 001000 |
| 9 | pedestrian crossing ahead | PCA | $P_1$ | 100/1075 | 110 |
| 10 | work in progress ahead | WPA | $P_3$ | 25/1075 | 001011 |
| 11 | leave way for the ambulance | LWA | $P_1$ | 100/1075 | 011 |
| 12 | intersection ahead | ISA | $P_2$ | 50/1075 | 00000 |
| 13 | taking left turn | TLT | $P_2$ | 50/1075 | 00011 |
| 14 | taking right turn | TRT | $P_2$ | 50/1075 | 00010 |
| 15 | road condition not good | RNG | $P_3$ | 25/1075 | 001010 |
| 16 | allow overtake | AWO | $P_3$ | 25/1075 | 010101 |
| 17 | allowed overtake | AEO | $P_3$ | 25/1075 | 010100 |
| 18 | searching for parking | SFP | $P_3$ | 25/1075 | 01011 |
| 19 | taking u turn | TUT | $P_2$ | 50/1075 | 01001 |
| 20 | vehicle turning in front | VTF | $P_2$ | 50/1075 | 01000 |

For the vehicle to vehicle communication based on the probability of occurrences of small case letters, and space in the English language as shown in Table I the Huffman tree is generated. To transmit the safety messages, each symbol of the words is mapped to the code-word based on the Huffman tree. These compressed assigned bits are sent to the receiver. The receiver then decodes the message using the Huffman tree known at the receiver. Thus, the words can be recovered at the receiver. The advantage of using a fixed tree is that the overhead of the tree's generation is reduced, transmitting the mapping of the symbols and their respective code-word to the receiver is reduced. Adaptive versions of Huffman coding have the following disadvantages they are prone to errors, the compression achieved is not significant, and each time for a new symbol, the tree has to be updated, which is computationally expensive for this application.

*B. Arithmetic coding*

Arithmetic coding [10] is an entropy-based lossless source coding technique. Arithmetic coding maps or encodes the entire message into a single number. For the long sequences with skewed distribution and a small number of symbols, arithmetic coding performs better than Huffman coding.

For encoding the messages for the vehicle to vehicle communication based on the distribution of small case letters and space in the English language as shown in Table I, the message is mapped to its corresponding tag using the implementation mentioned in [10]. The received tag is decoded at the receiver.

*C. Lempel-Ziv-Welch coding*

Lempel-Ziv-Welch (LZW) [11] is dictionary-based coding. It is a form of universal lossless source coding technique. For text data, the algorithm maps each character (sequence of 8-bit) data to the corresponding index of the dynamically growing dictionary index (fixed-length 12-bit codes). Dictionary indices from 0 to 255 represent single characters that are initialized at the start. The dictionary index from 256 to 4095 is created dynamically as the sequences in the data encountered during the encoding process. Lempel-Ziv-Welch algorithm performance is better if there is a repetition of the pattern in the message to be transmitted.

In our implementation, we have initialized the dictionary to contain small case English alphabets and space. In the original application, as mentioned in [11], the 8-bit data sequence is mapped to a fixed-length 12-bit code. A small modification is done. Instead of representing each character (series of 8-bit data) as a fixed-length 12-bit dictionary index, the first four bits are appended at starting the compressed sequence. Bits are added to indicate the fixed number of bits that are used to represent each of the dictionary index locations. This modification is helpful as it reduces the number of bits to be transmitted.

*D. Abbreviation based coding*

This method of coding is similar to Ham radio abbreviations. The safety messages are assigned abbreviations, as shown in the third column of Table II. Instead of transmitting the entire safety message, the abbreviation corresponding to the signal is sent. Thus the number of bits to be transmitted is reduced. The mapping of the word to its abbreviation is familiar to both the encoding and the decoding part. As shown in Table II, any safety message is represented by a fixed three-character abbreviation. Thus, any safety message can be transmitted using 24 bits.

*E. Probability-based coding*

The principle used in probability-based coding is the entire message is mapped to a code-word instead of coding each symbol of the signal to be transmitted as shown in Figure 3. A fewer number of bits are used to represent the priority-based or emergency based messages as they need to be sent as quickly as possible.

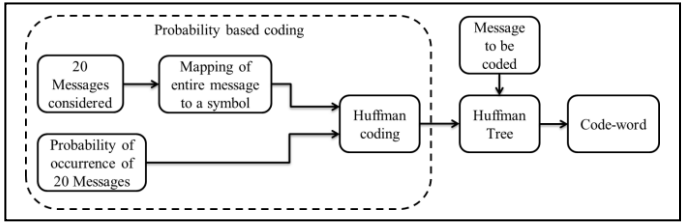

Fig. 3. Block diagram for probability-based coding

As the number of occurrences of these messages is not known, the words are classified into three priority levels $P_1$, $P_2$, and $P_3$, respectively, as shown in the fourth column of Table II. The letters with priority $P_1$ are the messages that occur frequently or are emergency messages and should be represented with fewer bits. Similarly, the message with priority $P_2$ is the messages that occur less often or not the emergency message. The messages with priority $P_3$ are the messages used for additional precautions or additional information so that they can be represented by comparably more number of bits. The probability of occurrence is assigned based on these priority levels for each message, as shown in the fifth column of Table II. Using Huffman coding, the code-word for each word is obtained, as shown in the sixth column of Table II. For example, the code-word 00111 is transmitted instead of transmitting left turn ahead. Thus, the number of bits to be transmitted is reduced. This method is suitable for sending essential safety messages for vehicle to vehicle communication.

## IV. RESULTS AND DISCUSSIONS

In this section, we compare the results obtained for the compression algorithms discussed in Section 3. The algorithms implementation was done on Windows 10, a 64-bit operating system using MATLAB Version 9.4 (R2018a). The algorithms are compared, considering the compression ratio and the time required for encoding and decoding. The compression ratio is obtained using the equation, as shown in Equation 1.

$$\text{Compression ratio} = \frac{\text{Uncompressed Size}}{\text{Compressed Size}} \quad (1)$$

Here, the second parameter considered for comparison is the time required for encoding and decoding. This parameter is necessary since it needs to be considered for efficient real-time encoding and decoding. We have used the MATLAB commands for computing the time required for encoding and decoding, namely *tic* and *toc*.

Figure 4 shows the comparison of the results for transmitting the priority $P_1$ message *leave way for the ambulance*. It can be observed that the Probability-based coding has a better compression ratio, as shown in Figure 4a. On the other hand, the time required for encoding and decoding is better for abbreviation-based coding, as shown in Figure 4b. Probability-based coding has a better compression ratio to the time required.

Similarly, Figure 5 compares the results for transmitting the priority $P_2$ message *left turn ahead*. It can be observed that Probability-based coding has a better compression ratio, as shown in Figure 5a. On the other hand, the time required for encoding and decoding is better for abbreviation-based coding, as shown in Figure 5b. Since we aim to obtain a better compression ratio with lesser processing time, probability-based coding has a better compression ratio to the time required.

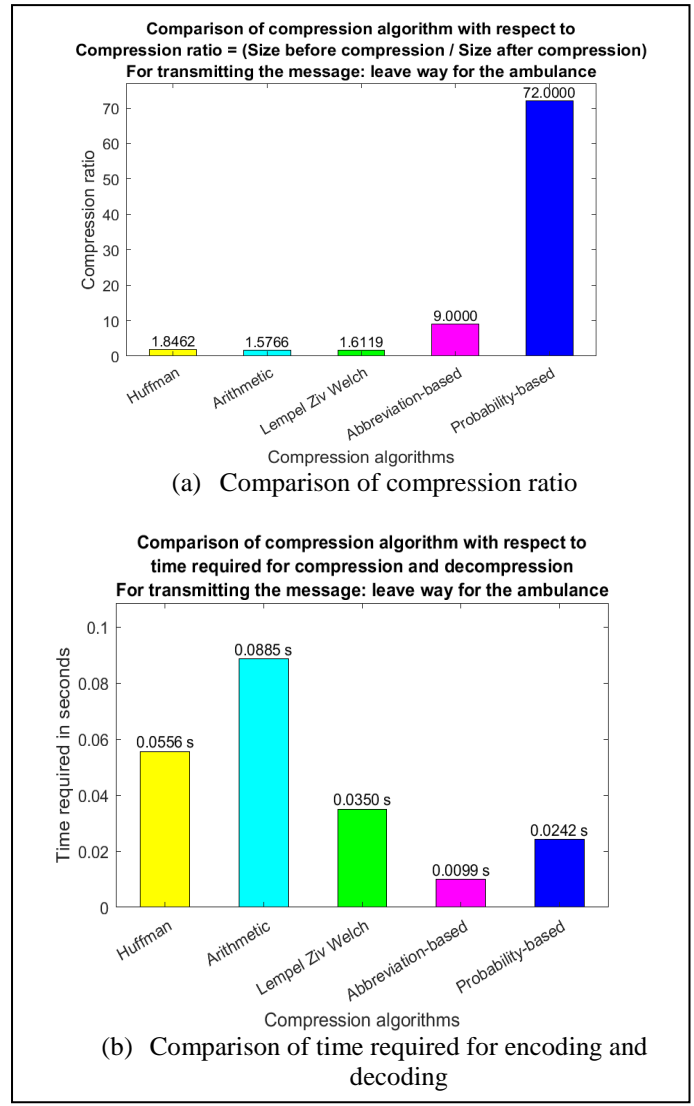

Fig. 4. Comparison of compression ratio and time required for transmitting the message leave way for the ambulance

Similarly, Figure 6 shows the comparison of the results for transmitting the priority $P_3$ message *road condition not*

*good*. It can be observed from Figure 6a and Figure 6b that the probability-based coding technique has a better value of compression ratio to the time required.

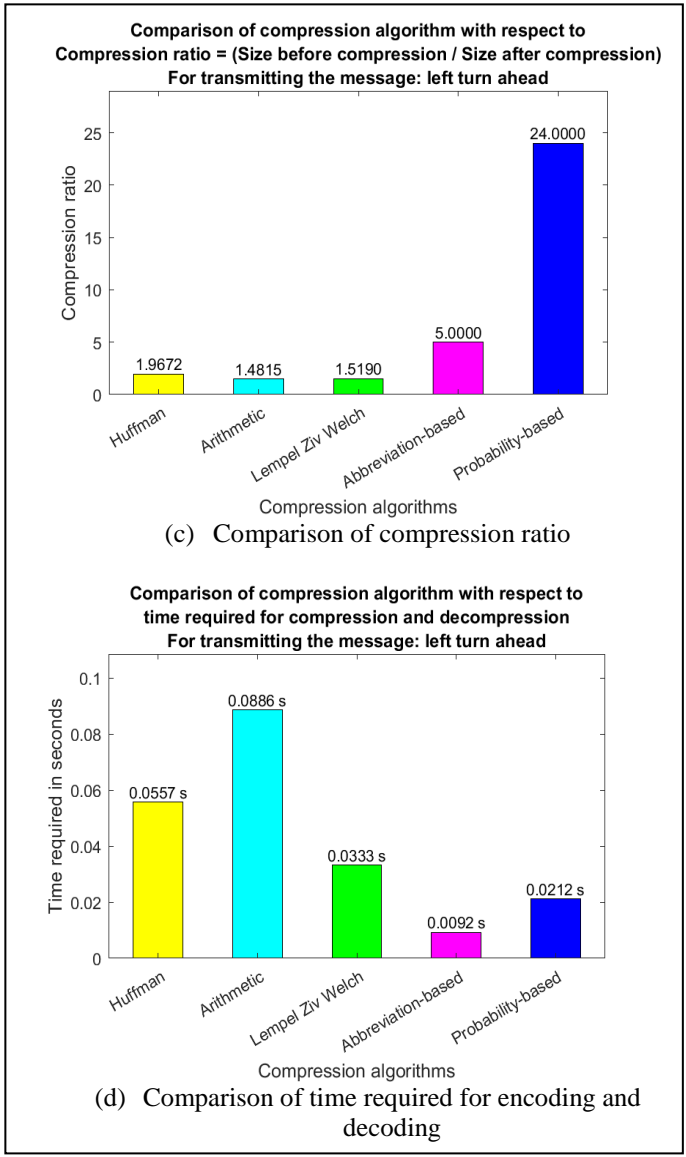

(c) Comparison of compression ratio

(d) Comparison of time required for encoding and decoding

Fig. 5. Comparison of compression ratio and time required for transmitting the message left turn ahead

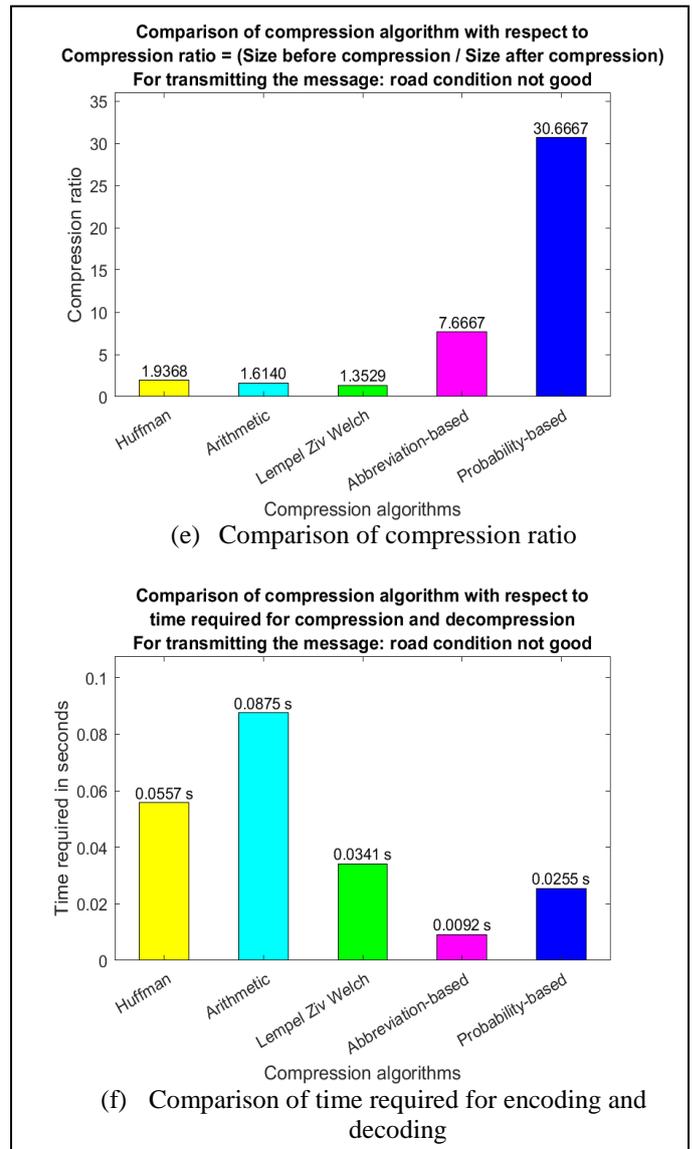

(e) Comparison of compression ratio

(f) Comparison of time required for encoding and decoding

Fig. 6. Comparison of compression ratio and time required for transmitting the message road condition not good

## V. CONCLUSIONS

In this paper, abbreviation-based and probability-based source coding techniques are proposed for the vehicle to vehicle communication for the considered safety messages. The paper compares the various algorithms for compression ratio and time required for encoding and decoding. The performance of probability-based coding is better by considering the proportion of the compression ratio to the time required.